\documentclass{article}
\usepackage{spconf,amsmath,graphicx}

\usepackage{float}
\usepackage{amsmath,graphicx,amsfonts,amssymb,url}
\usepackage{setspace}
\usepackage[caption=false]{subfig}
\usepackage{blindtext}
\usepackage[inline]{enumitem}
\usepackage{color}
 \usepackage{subfig}
\usepackage{multirow}
\newsavebox{\measurebox}
\usepackage{minibox}
\usepackage{array,multirow,booktabs}
\usepackage{graphicx}
\usepackage{siunitx}
\usepackage{amsfonts,amssymb}
\usepackage{balance}
\usepackage{pifont} 
\newcommand{\squeezeup}{\vspace{-2.5mm}}

\usepackage{algorithm}
\usepackage{algorithmic}
\usepackage{url}

\newcommand*{\affaddr}[1]{#1} 
\newcommand*{\affmark}[1][*]{\textsuperscript{#1}}

\usepackage{color}

\title{A Strong Baseline for Image and Video Quality Assessment}
\name{Shaoguo Wen\affmark[1], Junle Wang\affmark[1]}
\address{\affaddr{\affmark[1]Turing Lab, \  Tencent }}
 
\begin{document}
 
\maketitle

\begin{abstract}
In this work, we present a simple yet effective unified model for perceptual quality assessment of image and video. In contrast to existing models which usually consist of complex network architecture, or rely on the concatenation of multiple branches of features, our model achieves a comparable performance by applying only one global feature derived from a backbone network (i.e. resnet18 in the presented work). Combined with some training tricks, the proposed model surpasses the current baselines of SOTA models on public and private datasets. Based on the architecture proposed, we release the models well trained for three common real-world scenarios: UGC videos in the wild, PGC videos with compression, Game videos with compression. These three pre-trained models can be directly applied for quality assessment, or be further fine-tuned for more customized usages. All the code, SDK, and the pre-trained weights of the proposed models are publicly available at https://github.com/Tencent/CenseoQoE.

\end{abstract}
\section{Introduction}
\label{sec:intro}
Image/Video quality assessment(I/VQA) have been a long-standing problem in image/video processing and computer vision, always used as a measurement or optimization target in the fields of video compression, quality monitoring, video recommendation systems and etc. Nowadays, user-generated content (UGC) and video streaming has exploded on the Internet, the enormous amount of video storage and transmission poses new challenges to the size of the video, I/VQA can provide measurement for encoders to reduce the bit rate of the video or compression algorithms to compress the video with little or no perceptual impact on the video quality. Another novel application is used in the recommendation system to provide users with higher quality videos. Taking advantage of such optimizations allows for better user experience at lower cost for the provider which shows great value.

Generally, quality assessment can be categorized into subjective assessment and objective assessment. Subjective assessment usually requires a certain number(15 at least according to ITU-R BT.500 \cite{bt2002methodology} ) of people to evaluate the quality of image or video, then mean of opinions(MoS) is regarded as the final quality score. Subjective assessment always obtain reliable and accurate results for quality assessment, however, it is too expensive and time-consuming to be used in the quality evaluation of visual systems that requires frequent and real-time feedback. The objective assessment predicts a quality score by algorithms that aims to correlate well with human perception which is significantly more piratical for real-time image/video quality evaluation. 

Many efforts have been made on developing objective algorithms of image/video assessment.The Peak Signal to Noise Ratio (PSNR)~\cite{wang2009mean}, the Structural Similarity Index (SSIM)~\cite{wang2004image} and the Multi-Scale Structural Similarity (MS-SSIM)~\cite{wang2003multiscale} are usually used as traditional methods for image assessment, but they are not correlate well with human perceptual quality sometimes. The Video Multi-Method Assessment Fusion (VMAF)~\cite{li2018vmaf} take use of hand-crafted features and machine learning to generate model to predict quality of videos, but which is limited to when the reference video is available. Deep learning have achieved great success on computer vision in recent years, many works\cite{liu2017rankiqa}\cite{tu2021rapique}\cite{ying2021patch}\cite{wang2021rich} applied CNN and RNN to tackle the problem of I/VQA and achieve high performance. However, we found that many previous works were expanded on poor baselines, besides, the comparison between methods is unfair because some of them obtained the improvement by training with tricks rather than proposed methods themselves. In addition, many state-of-the-arts (SOTA) models designed complicated network architecture which is not suitable for industrial deployment. In this paper, we proposed a simple and effective unified model for image/video quality assessment which acquires a strong baseline by training with some common tricks.

To demonstrate the performance of our method, we conduct experiments on three publicly available databases, i.e., LIVE-VQC \cite{sinno2018large}, KoNViD-1K\cite{hosu2017konstanz} and YouTube-UGC\cite{wang2019youtube}, and three private datasets in different commercial scenarios, i.e., UGC videos in the wild, PGC videos with compression and Game videos with compression.

The main contributions of this work are as follows:
\begin{itemize}
\item[$\bullet$] For the academia, we hope the strong baseline provided by our proposed method help researchers to design more excellent models and achieve higher performance in the I/VQA community.
\item[$\bullet$] For the industry, the model we proposed is simple but high in performance without extra inference consumption, which is useful for industrial deployment of I/VQA models to achieve the goal of real-time feedback.
\item[$\bullet$] We release three model weights of our proposed method that have been trained on three carefully designed datasets of different real-world commercial scenarios, which can be directly used for quality assessment or fine tuned on own dataset. 
\end{itemize}

\section{Related Work}
\label{sec:RW}\squeezeup
\subsection{Image Quality Assessment} 
Image Quality Assessment(IQA) can be classified into distortion-specific methods and general-purpose methods according to \cite{zhu2020metaiqa}. The distortion-specific methods\cite{li2015no}\cite{li2013referenceless} evaluated the image quality by extracting features of known distortion types, but their application scope is limited because the distortion types are always unknown or mixture. The general-purpose methods are further divided into Natural Scene Statistics (NSS) methods and learning-based methods. The NSS methods extracted features in different sub-bands and estimate the distributional parameters for predicting quality. In learning-based methods\cite{chetouani2010novel}\cite{ye2012no}\cite{ye2012unsupervised}, features are extracted and mapped to the MOS by Support Machine Regression or Neural Networks. Deep learning based methods have been developed by many works in recent years which resulted in significant improvements and showed great potential. Kang et al.\cite{kang2014convolutional} applied CNN to train IQA model 
with small image patches rather than images, which improved the performance of the model by augmenting training examples. Liu et al.\cite{liu2017rankiqa}\cite{liu2019exploiting} combined CNN with ranking learning to further improve the performance of models. Hossein Talebi and Peyman Milanfar\cite{talebi2018nima} proposed a novel approach to predict both technical and aesthetic qualities of image. Zhu et al.\cite{zhu2020metaiqa} proposed a no-reference IQA metric based on deep meta-learning which tried to learn the meta-knowledge shared by human when evaluating the quality of images with various distortions.
 
\subsection{Video Quality Assessment} 
Most early VQA models were distortion specific \cite{argyropoulos2011no}\cite{pandremmenou2015no}\cite{vega2017predictive} and focused mostly on transmission and compression related artifacts. Li et al. proposed a learning-based method for FR-VQA named Video Multi-Method Assessment Fusion (VMAF)~\cite{li2018vmaf} which extracts features from videos and trains a Support Vector Machine(SVM) model to predict quality of videos. Similiar with IQA, deep learning-based methods obtained promising results in VQA in recent years, Kim et al. \cite{kim2018deep} utilize CNN models to learn the spatial-temporal sensitivity maps. Liu et al. \cite{liu2018end} exploit a 3D-CNN model for codec classification and quality assessment of compressed videos. Wang et al.\cite{wang2021rich} create a large scale UGC video dataset and propose a DNN-based framework to thoroughly analyze importance of content, technical quality and compression level in perceptual quality. Tu et al. \cite{tu2021rapique}  proposed an efficient model for predicting the subjective quality of UGC videos which leverages a composite of spatio-temporal scene statistics features and deep CNN-based high-level features. Ying et al. \cite{ying2021patch} created a largest(by far) in the wild UGC video quality dataset and proposed two unique NR-VQA models: a local-to-global region-based NR VQA architecture and a first-of-a-kind space-time video quality mapping engine.

\section{The Proposed Objective Model}
\label{sec:Obj} \squeezeup
\subsection{Network Architecture} 
Conventionally, image/video quality assessment(IVQA) can be divided into three main categories:full-reference (FR), reduced-reference (RR), and no-reference (NR) models. FR model predicts quality score against pristine image/video, while no-reference (NR) model involve no such comparison. 

\begin{figure}[!htbp]
    \centering
    \includegraphics[width=\columnwidth]{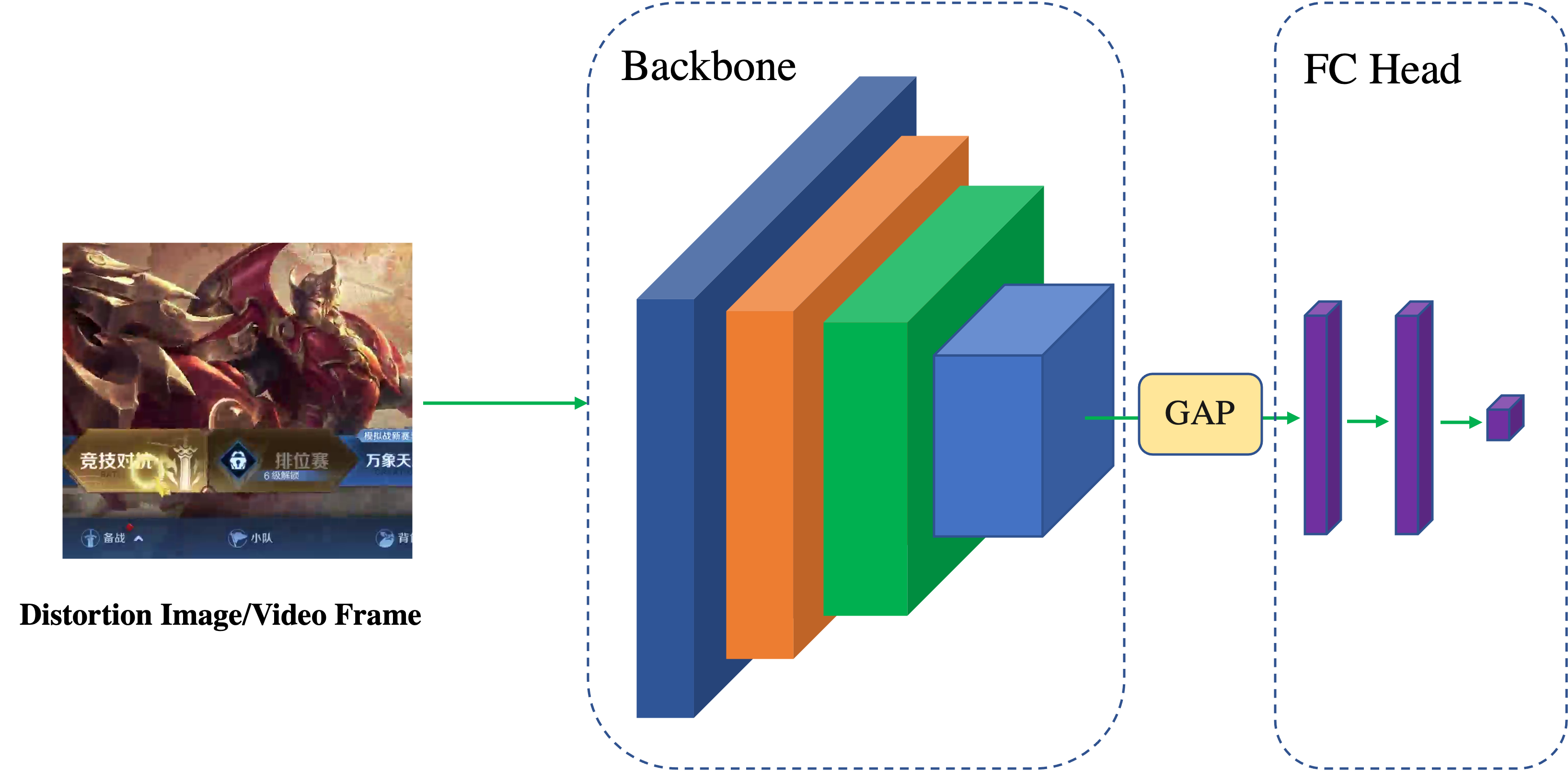}
    \caption{ Network Architecture of NR model.}
     \squeezeup  
    \label{fig:arch_nr}
\end{figure}

\begin{figure}[!htbp]
    \centering
    \includegraphics[width=\columnwidth]{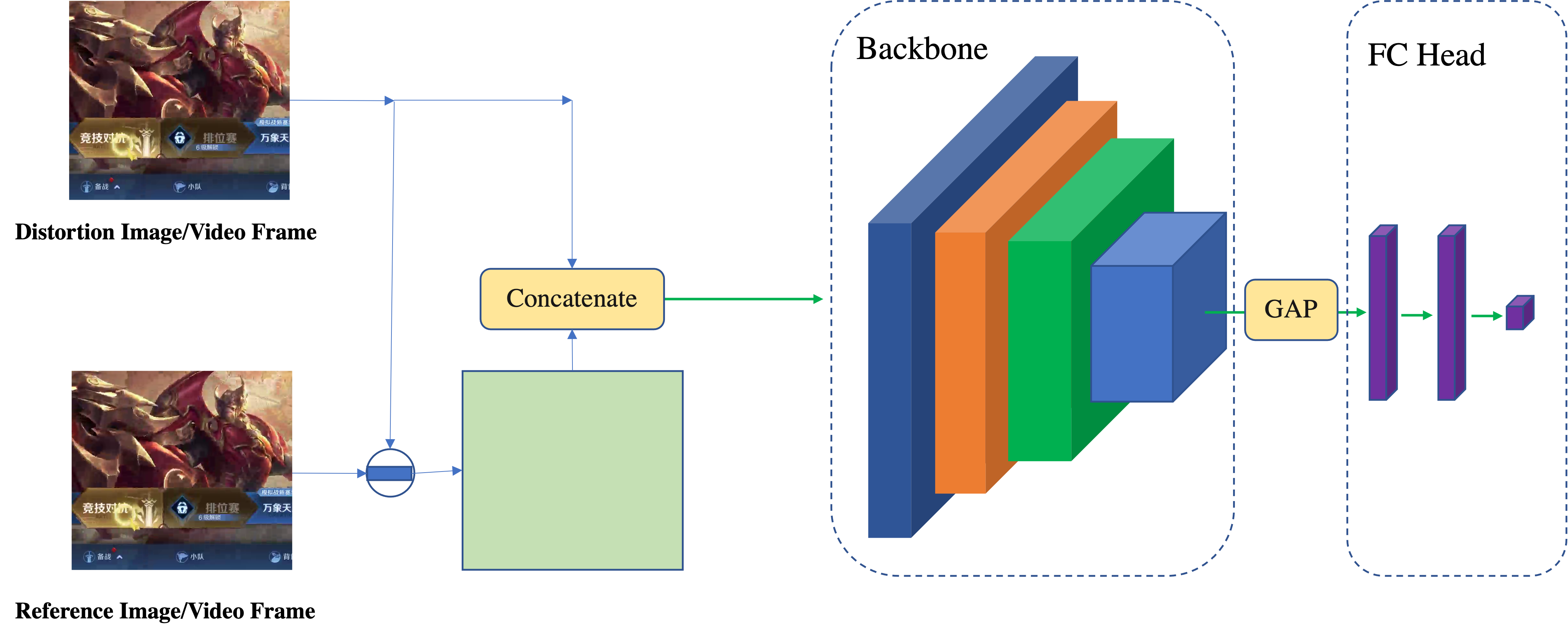}
    \caption{ Network Architecture of FR model.}
     \squeezeup  
    \label{fig:arch_fr}
\end{figure}

A simple but effective network architecture is proposed for FR and NR models in this paper as depicted in Fig.~\ref{fig:arch_nr} and Fig.~\ref{fig:arch_fr} respectively. A light-weight network is applied as the backbone of proposed model for efficient inference, such as Mobilenet~\cite{howard2017mobilenets}, Shufflenet~\cite{zhang2018shufflenet}, ResNet-18\cite{he2016deep}~\textsl{etc.} The output feature of last convolution layer in the backbone is fed into the Global Averaged Pooling(GAP) ~\cite{szegedy2015going} module, two Fully Connected (FC) layers with 1024 hidden nodes take the flatten feature obtained by GAP as inputs and predict the final quality score. For image quality assessment, image to be evaluated is fed into the network directly and obtain the quality score. For video quality assessment, video is first extracted and each frame is fed into the model, and then the scores of all frames are frames-wise averaged to obtain the final quality score. The main difference between our proposed NR and FR models is in the input of the network, the NR model take the distortion image as input directly, however, in the FR model, the reference image is first subtracted by the distortion image, the result of subtraction is concatenated with the distortion image and then fed into the FR network. Note that the dimension of first convolution layer should modified from 3 to 6 due to the concatenation in the FR model. In our experiment, the backbone of our model is pre-trained on Imagenet and fine tuned on quality data.

\subsection{Loss function}
We denote $\hat{y}$ as the predicted score by the objective model, and let $y$ be the ground truth quality score collected from the subjective experiment. $n$ is the batch size of input images in the training phase. The loss function of our proposed model is composed of two parts as defined below, where $ \lambda $ is a parameter that balances the two losses:
\begin{equation}
L = L_{mae} + \lambda \cdot L_{rank}.
\label{eq:overall}
\end{equation}

The first part is the Mean Absolute Error (MAE) loss $L_{mae}$ between the ground truth and predicted scores:
\begin{equation}
L_{mae} = \frac{1}{n}\sum^n_{i=1} | \hat{y_i} - y_i |.
\end{equation}

The second part is a pair-wise ranking loss $L^{ij}_{rank}$ which is inspired by the metric learning of image quality assessment proposed in~\cite{liu2017rankiqa}. Different with ~\cite{liu2017rankiqa}, instead of synthetically generating deformations of images over a range of distortion intensities, we apply rank learning in the training data. Specifically,  given arbitrary pair of images in the batch inputs, the proposed $L^{ij}_{hard}$ is designed as:
\begin{equation}
L^{ij}_{rank} = max (0, |y_i-y_j|-e(y_i, y_j)\cdot (\hat{y_i} -  \hat{y_j})),
\end{equation}

where $e(y_i, y_j)$ is defined as:
\begin{equation}
e(y_i, y_j) = \left\{ \begin{array}{rcl}
 1, \ \ \ \ \ \  y_i \geq y_j  \\  
 -1,  \ \ \  \ \ \  otherwise \ \ \ 
\end{array}\right.
\label{eq:l1}
\end{equation}

Finally, $L_{rank}$ is calculated by:
\begin{equation}
L_{rank} = \frac{1}{n\cdot n}\sum^n_{i=1}\sum^n_{j=1} L^{ij}_{rank}
\end{equation}

$L_{rank}$ help model capture more detailed information over different distortion or different degrees under same distortion, besides, which can also  speed up the convergence of the model.

\squeezeup
\subsection{Training Tricks}
\label{sec:tt} \squeezeup
\label{sec:md}
Cosine annealing learning rate decay proposed in SGDR\cite{loshchilov2016sgdr} is applied as learning rate schedule in our training phase, only the cosine annealing part is implemented without the restarts part. Supposed that $T$ is the max training epochs, $l_{init}$ is the initial learning rate at the beginning of training, and  $l_{min}$ is the minimum learning rate at the end of training, then the decaying of learning rate over training is denoted as:

\begin{equation}
L(t) =  l_{min} + \frac{1}{2}(l_{init}-l_{min})(1+cos(\frac{\pi*t}{T})
\end{equation}
Where $t$ is the current epoch.

Unlike common computer vision tasks, whose inputs can be resized as any shape for fitting the input shape of model, but in the image/video quality assessment tasks, images should be resized with same ratio in avoid to introduce unnecessary distortion and mislead the training of model. According to the setting of Quality of Experiment, we resize short size of image to the max resolution of all quality data, e.g. 1080p, then random cropping is employed on the resized image to match the input shape of model.  Resizing and random cropping is a kind of data augmentation which can improve the performance of model. Note that center cropping is applied in the test phase.

The Adam and SGD with momentum optimizer are both implemented in our experiment. Empirically, Adam optimizer is suitable for training from scratch or pre-trained on Imagenet, which can accelerate the convergence of model. SGD optimizer with momentum is suitable for training with pre-trained on quality data, which can improve the robustness of model, especially when evaluating across datasets.

Stochastic Weight Averaging (SWA) \cite{izmailov2018averaging} is optionally implemented in our experiment, the key idea of SWA is to average multiple model weights produced by SGD with a modified learning rate schedule, which can reach a wider optima for better generalization. Note that SWA is not plugged in when compared with other methods. SWA usually obtains a better generalization without increasing the complexity of the model.

\squeezeup
\section{Experiment}
\label{sec:Ex} \squeezeup
\subsection{Experimental setup}
During training, in addition to the resizing and random cropping introduced in section ~\ref{sec:tt}, randomly flipped left to right is further implemented for data augmentation. The initial learning rate $l_{init}$ and minimum learning rate $l_{min}$ are set as $10^{-04}$ and $10^{-07}$ respectively in the Cosine Annealing learning rate decay strategy. To avoid over-fitting, weight decay was set as $5^{-04}$ and the momentum is set to 0.9 if SGD optimizer is applied. $\lambda$ in equation (\ref{eq:overall}) was set to 1.  

\begin{table*}[!htbp]
\begin{center}
 \squeezeup  \squeezeup 
\caption{\label{tab:main performance}%
Performances comparison on public datasets. }
{
\renewcommand{\baselinestretch}{1}\footnotesize
\begin{tabular}{|c|c|c|c|c|c|c|} \hline
 &\multicolumn{2}{|c|}{LIVE-VQC} &\multicolumn{2}{|c|}{KoNViD-1K} &\multicolumn{2}{|c|}{YouTube-UGC}   \\ \hline
 & PLCC & SRCC & PLCC & SRCC  & PLCC & SRCC \\ \hline
 VGG-19\cite{tu2021ugc} & 0.7160 & 0.6568 & 0.7845 & 0.7741 & 0.6997 & 0.7025 \\ \hline
 ResNet-50\cite{tu2021ugc} & 0.7205 & 0.6636 & 0.8104 & 0.8018 & 0.7097 & 0.7183 \\ \hline
 RAPIQUE\cite{tu2021rapique} & \bf{0.7863} & \bf{0.7548} & 0.8175 & 0.8031 & 0.7684 & \underline{0.7591} \\ \hline
 PatchVQ\cite{ying2021patch} & 0.7205 & 0.6636 & \bf{0.837} & \bf{0.827} & - & - \\ \hline
 CoINVQ\cite{wang2021rich} & - & - & 0.767 & 0.764 & \bf{0.802} & \bf{0.816} \\ \hline
 \bf{Ours} & \underline{0.7575} & \underline{0.7390} & \underline{0.8245} & \underline{0.8185} & \underline{0.7691} & 0.7554 \\ \hline
\end{tabular}}
\end{center}
\squeezeup  \squeezeup
\label{tab:res}
\end{table*}

We conduct evaluation experiment of our proposed model on the three public UGC-VQA databases: LIVE-VQC \cite{sinno2018large}, KoNViD-1K\cite{hosu2017konstanz} and YouTube-UGC\cite{wang2019youtube}. All the datasets are randomly split into non-overlapping training and test sets (80\%/20\%), this process of random split was repeat 20 times and the overall median performance was recorded. It’s also worth noting that different videos belonging to the same reference video should be contained in the same train set or test set. Besides, we also evaluate our model on three private datasets in  different commercial scenarios: UGC videos in the wild, PGC videos with compression and Game videos with compression, the results demonstrates the effective and high performances of our model. To evaluate the performances of model, the Pearson Linear Correlation Coefficient (PLCC) and  Spearman’s rank order correlation coefficient (SRCC) are considered as evaluation metrics.
 
\subsection{Experimental results}
The results on public datasets are shown in Table~\ref{tab:main performance}, where the best and second-best results are respectively marked in bold and underlined fonts. As shown in the table, our model obtain second-best performance of PLCC and SRCC in LIVE-VQC and KoNViD-1K datasets and second-best performance of PLCC in YouTube-UGC. The gap of performance between SOTA models and our model is not very large but our model has fewer parameters and faster inference speed. Besides, the performance of our proposed model far exceeds those models that are often used as baselines, e.g. VGG-19 and ResNet-50. Our strong baseline can achieve 0.7575 in PLCC and 0.7390 in SRCC in LIVE-VQC dataset, which beats standard baseline VGG-19\cite{tu2021ugc} by more than 0.04 in PLCC and 0.08 in SRCC.

To evaluate the performance of our model in real commercial scenario, we conduct subjective experiment under the ITU-R BT.500 \cite{bt2002methodology} standard and build three datasets for training model according to our commercial scenarios, i.e. UGC videos in the wild, PGC videos with compression and Games videos with compression. We first collected and processed a certain number of videos in three commercial scenario data, specifically, around 3000 videos with compression for PGC and Games videos respectively and more than 20000 videos in the wild for UGC. During the experiment, 20-25 were asked to score the videos. The observers was asked to launch a internally  developed  platform to start the test, using their own mobile phone. In another word, different models of mobile phone were utilized by different participants to conduct the subjective test, which is consistent with real application scenario.
The performance of our model is shown on Table~\ref{tab:performance on private datasets}, our model achieves high PLCC and SRCC in all three private datasets. Unfortunately, these datasets are not yet publicly available due to some reasons, but models trained on these datasets already available at https://github.com/Tencent/CenseoQoE .

\begin{table}[!htbp]
\begin{center}
 \squeezeup  \squeezeup 
\caption{\label{tab:performance on private datasets}%
Performances on private datasets. }
{
\renewcommand{\baselinestretch}{1}\footnotesize
\begin{tabular}{|c|c|c|} \hline
 & PLCC & SRCC \\ \hline
 Games videos(compression) & 0.971 & 0.968 \\ \hline
 PGC videos(compression)& 0.961 & 0.959 \\ \hline
 UGC videos(in the wild) & 0.902 & 0.880 \\ \hline
\end{tabular}}
\end{center}
\squeezeup  \squeezeup
\label{tab:res}
\end{table}

\section{Conclusion} \squeezeup   
\label{sec:Con}
In this study, we propose an efficient and high-performance unified model for image/video quality assessment, which achieved comparable or even surpassing performance in some datasets compared with state-of-arts models and obtained a strong baseline in many public I/VQA datasets. Our proposed model achieves a high trade-off between performance and complexity, so it is very suitable for industrial deployment. Besides, three trained model based on our proposed method are released, which can be directly used for quality assessment in the real-world commercial scenario or fine tuned on own dataset.

\balance 
\bibliographystyle{IEEEbib}
\scriptsize{
\bibliography{refs}}

\end{document}